\documentclass{jfm}

\usepackage{graphicx}
\usepackage{float}
\usepackage{newtxtext}
\usepackage{amssymb}
\usepackage{color}
\usepackage{newtxmath}
\usepackage{natbib}
\usepackage{hyperref}
\usepackage{epsfig,subfigure}
\usepackage{amsmath}
\usepackage{amssymb}
\usepackage[utf8]{inputenc}
\graphicspath{{figures/}}
\usepackage{cleveref}
\crefname{section}{§}{§§}
\Crefname{section}{§}{§§}

\usepackage{float}
\hypersetup{
    colorlinks = true,
    urlcolor   = blue,
    citecolor  = black,
}

\newcommand{\RomanNumeralCaps}[1]

\title{Existence of bistable states in curved compression ramp flows}

\author{Yan-Chao Hu\aff{1,2}
	Gang Wang\aff{1,2},
	Wen-Feng Zhou\aff{1,2},
	Ming-Zhi Tang\aff{1,2},
	Ming Fang\aff{1,2},
	Yan-Guang Yang\aff{2,3}\corresp{\email{yangyanguang@cardc.cn}},
	\and Zhi-Gong Tang\aff{3}\corresp{\email{tangzhigong@126.com}}
}

\affiliation{\aff{1}Hypervelocity Aerodynamics Institute (HAI), China Aerodynamics Research and Development Centre, Mianyang 621000, China
	\aff{2}Laboratory of Aerodynamics in Multiple Flow Regimes (LAMFR), China Aerodynamics Research and Development Centre, Mianyang 621000, China
	\aff{3}China Aerodynamics Research and Development Centre, Mianyang 621000, China}
\begin{document}
	\maketitle
	
\begin{abstract}
\par This paper reports the bistability of curved compression ramp (CCR) flows. It reveals that both separation and attachment states can be stably established even for the same boundary conditions. Firstly, to investigate the effects of initial condition and evolutionary history, a thought experiment involving two processes with the same three steps but in different orders is designed, possibly constructing the two distinct-different stable CCR flows. Subsequently, three-dimensional direct numerical simulations are then performed to replicate the thought experiment, verifying the existence of the bistable states in CCR flows. In the end, the method of virtual separation disturbance is proposed to detect whether potential bistable states exist, theoretically demonstrating the bistability of CCR flows. As a canonical type of Shock wave-Boundary layer interactions, local CCR flows often appear on aircraft, hence the bistability will certainly bring noteworthy changes to the global aerothermodynamic characteristics, which supersonic/hypersonic flight has to deal with.
\end{abstract}


\section{Introduction}
\label{sec:Introduction}
\par In shock wave-boundary layer interaction (SBLI) flows, separation and attachment of boundary layers (BLs) are the two most typical states \citep{babinsky2011shock,chapman1958investigation}. The ultimate stable states of a specific SBLI flow are widely considered to be unique for given inflow parameters and wall geometry, i.e., the boundary conditions (BCs). Although empirical, this understanding is still correct in most cases. However, effects of initial condition and evolutionary history of SBLIs can not be ignored in principle, especially considering the reciprocal causation of shock wave (SW) patterns and BL behaviors. An extreme question stemming from those process-dependent effects is that "can both stable separation and attachment states exist for the same BCs in pure SBLI flows?" The statement "pure" means that the flow is dominated by self-organized SBLIs rather than other multistable interactions, such as SW reflections \citep{ben2001hysteresis,tao2014viscous}. If the answer is yes, the corresponding bistability will certainly bring significant differences to the aerodynamic characteristics on the wall, which is inevitable for super/hypersonic flight in multiple flow regimes. Herein, a class of pure SBLI flows, the canonical compression ramp (CR) flows, are chosen to investigate those effects, and the bistability of CR flows is reported.
\par CR flows's stable separation (SS) states are composed of large separation bubbles and $\lambda$-shock patterns classified as type VI by \citet{edney1968anomalous}. As a basic wall geometry, direct CR (DCR, the inclined ramp connecting the flat plate directly with one apex $O$) inducing SBLIs are investigated widely, including the interaction processes with BLs' distortions near separation \citep{chapman1958investigation,stewartson1969self,neiland1969theory}, the aerodynamic characteristics \citep{gumand1959on,hung1973interference,simeonides1995experimental,tang2021aerothermodynamic}, the vortex structures inside both the separation bubbles \citep{gai2019hypersonic,cao2021unsteady} and the evolving BLs \citep{fu2021shock,hu2017beta}, the unsteadiness of the SWs and BLs \citep{ganapathisubramani2009low,helm2021characterization,cao2021unsteady}. On the other hand, due to having the potential to weaken separation, curved CR (CCR, inclined ramp connecting the flat plate with curved walls) flows have gradually attracted the attentions of researchers \citep{tong2017direct,wang2019amplification,hu2020bistable}. However, as far as we know, there are few reports about the multistability of pure SBLIs.
\par The rest of this paper is organized as follows. Sec. \ref{sec:Thought experiment} describes a thought experiment to construct two distinct-different stable CCR flows for the same BCs. Sec. \ref{sec:Numerical expriment} presents the details of the three-dimensional (3D) direct numerical simulations (DNSs) to replicate the thought experiment, verifying the existence of bistable states in the CCR flows. Sec. \ref{sec:Demonstration} gives the theoretical proof of CCR flows' bistability to reveal its rationality. Conclusions follow in Sec. \ref{sec:Conclusions}. Some relevant movies are also given as supplementary materials.
\section{Thought experiment}
\label{sec:Thought experiment}
In this section, we design a thought experiment to obtain two possible distinct-different stable CCR flows, SS and stable attachment (SA), via processes 1 and 2, respectively. 
\par First, we construct a possible SS CCR flow via process 1 using the following three steps.
\begin{figure}
	\centering
	\subfigure[\label{subfig:Process1_Step1}]{\includegraphics[width = 0.32\columnwidth]{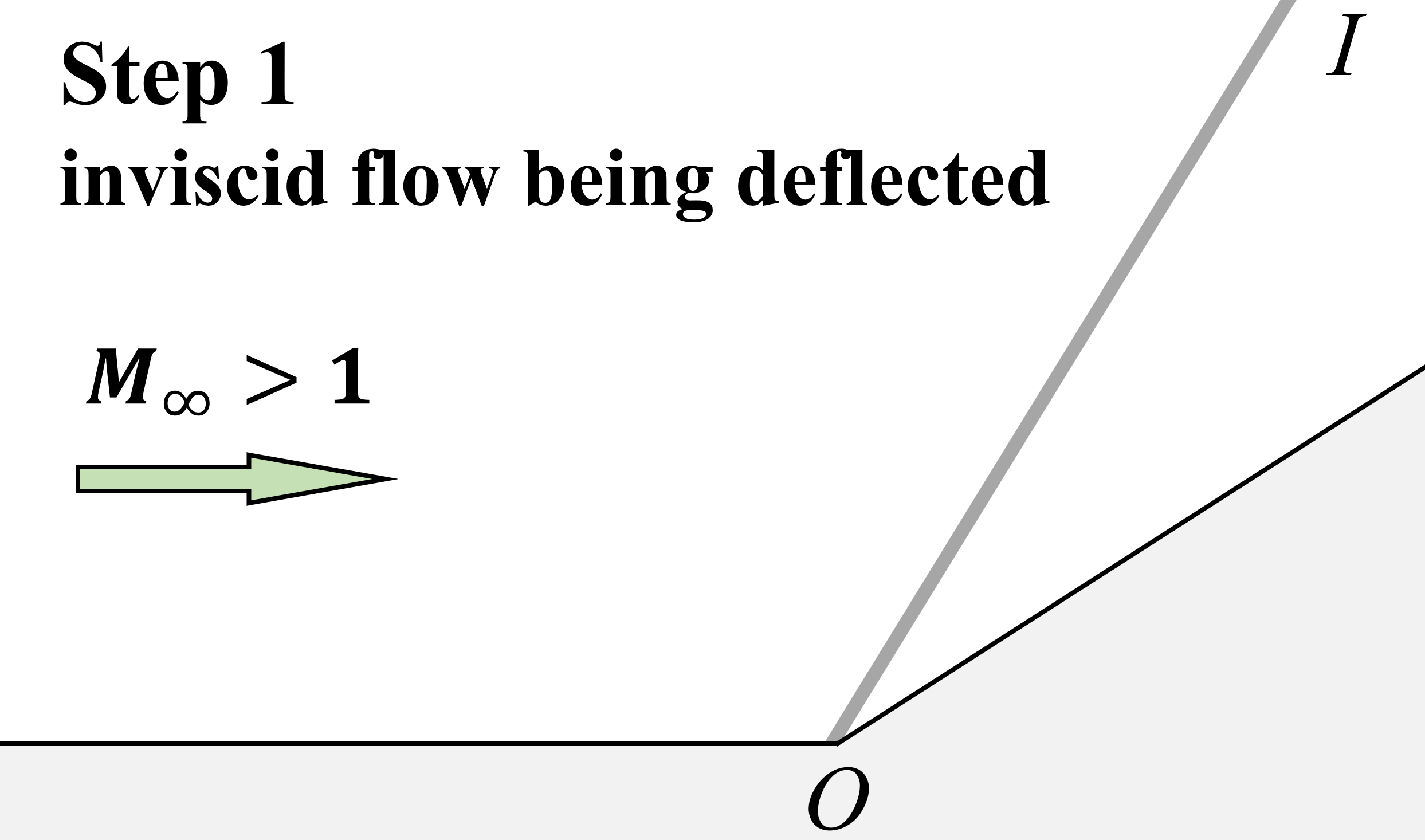}}
	\subfigure[\label{subfig:Process1_Step2}]{\includegraphics[width = 0.32\columnwidth]{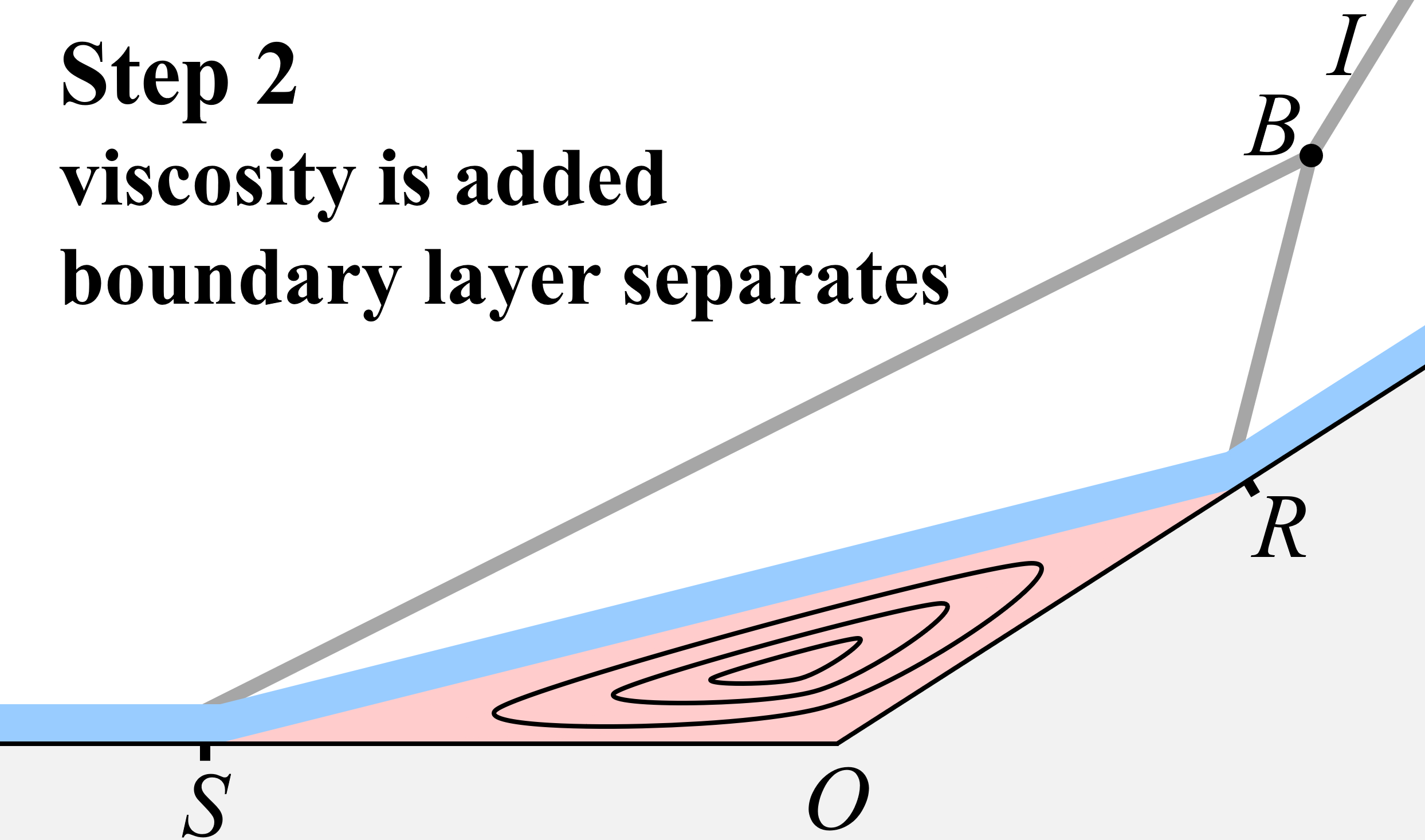}}
	\subfigure[\label{subfig:Process1_Step3}]{\includegraphics[width = 0.32\columnwidth]{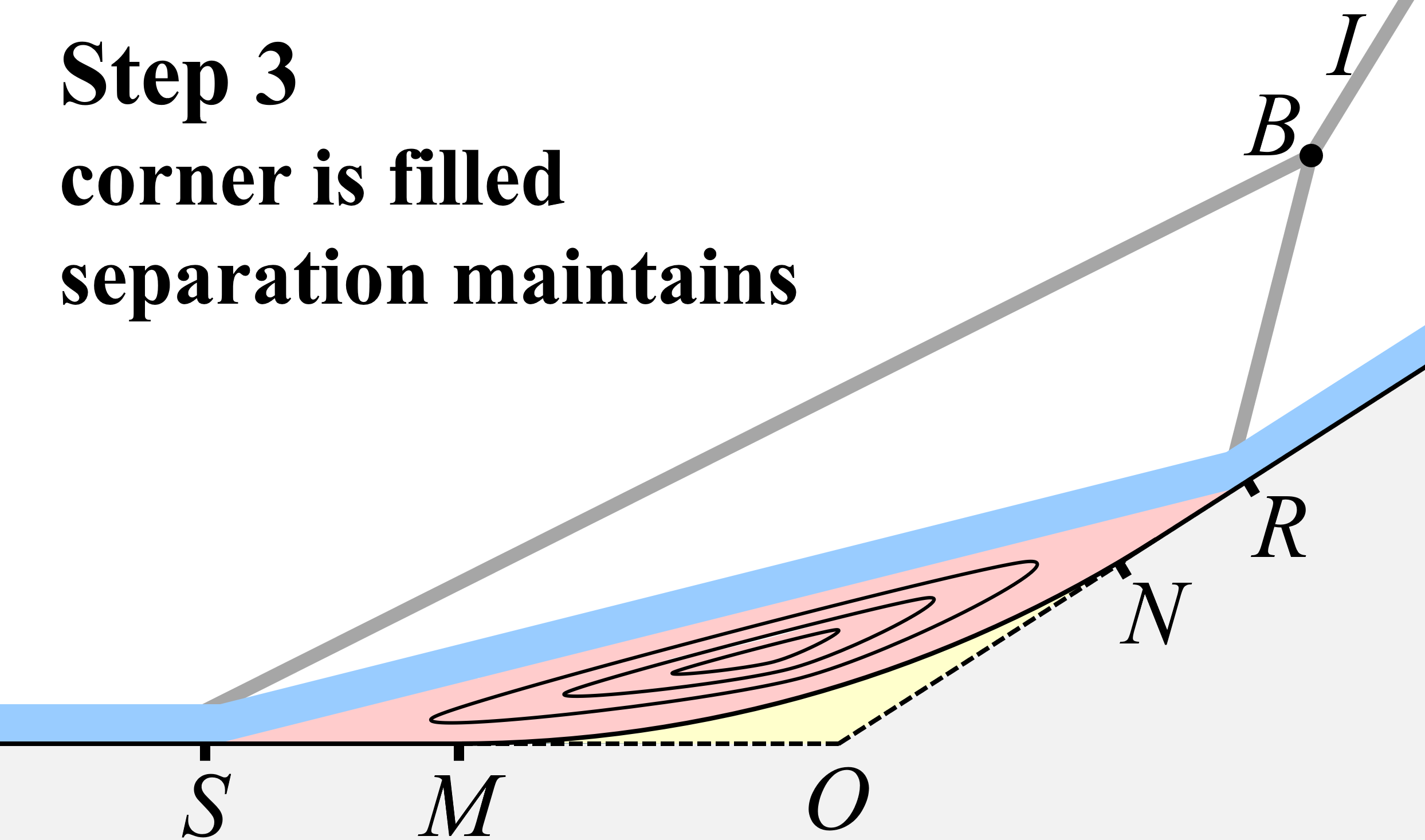}}
	\caption{Process 1 to construct a separated CCR flow. (a)Step 1;(b)step 2;(c)step 3.}
\end{figure}
\begin{itemize}
	\item {\bf Step 1: deflect the inviscid flow.} Set the initial flow field as an inviscid supersonic flow being deflected by a DCR geometry with ramp angle $\alpha$. An oblique SW $OI$ is formed and emits from the ramp apex $O$, as shown in Fig. \ref{subfig:Process1_Step1};
	\item {\bf Step 2: add the viscosity.} At some instant, the viscosity is suddenly added to the fluid. An attached BL (ABL) is subsequently formed and 
	interacts with SW $OI$. For a large enough $\alpha$, the ABL near point $O$ can not resist the strong adverse pressure gradient induced by $OI$ and then separates from the wall. The inverse flow gradually shapes a closed separation bubble $SRO$, deflecting both the ABL and the outer inviscid flow, and inducing two SWs $SB$ and $RB$. As time goes on, the separation bubble stops growing and stabilizes, i.e., both the separation point $S$ and the reattachment point $R$ become almost fixed or only oscillate slightly. The flow field is schematically shown in Fig. \ref{subfig:Process1_Step2};
	\item {\bf Step 3: fill the corner as an arc.} Fill the corner $O$ slowly and gently with an `imaginary machine' that can produce solid wall material atom by atom. In order to minimize the disturbance, do not fill the next atom till the flow stabilizes at a new SS state each time. In the end, the corner $O$ is filled as an arc $\overline{MN}$ inside space $SRO$, and an SS CCR flow could be obtained, as shown in Fig. \ref{subfig:Process1_Step3}.
\end{itemize}

\par Second, we construct a possible SA CCR flow via process 2 using the following steps.
\begin{figure}
	\centering
	\subfigure[\label{subfig:Process2_Step1}]{\includegraphics[width = 0.32\columnwidth]{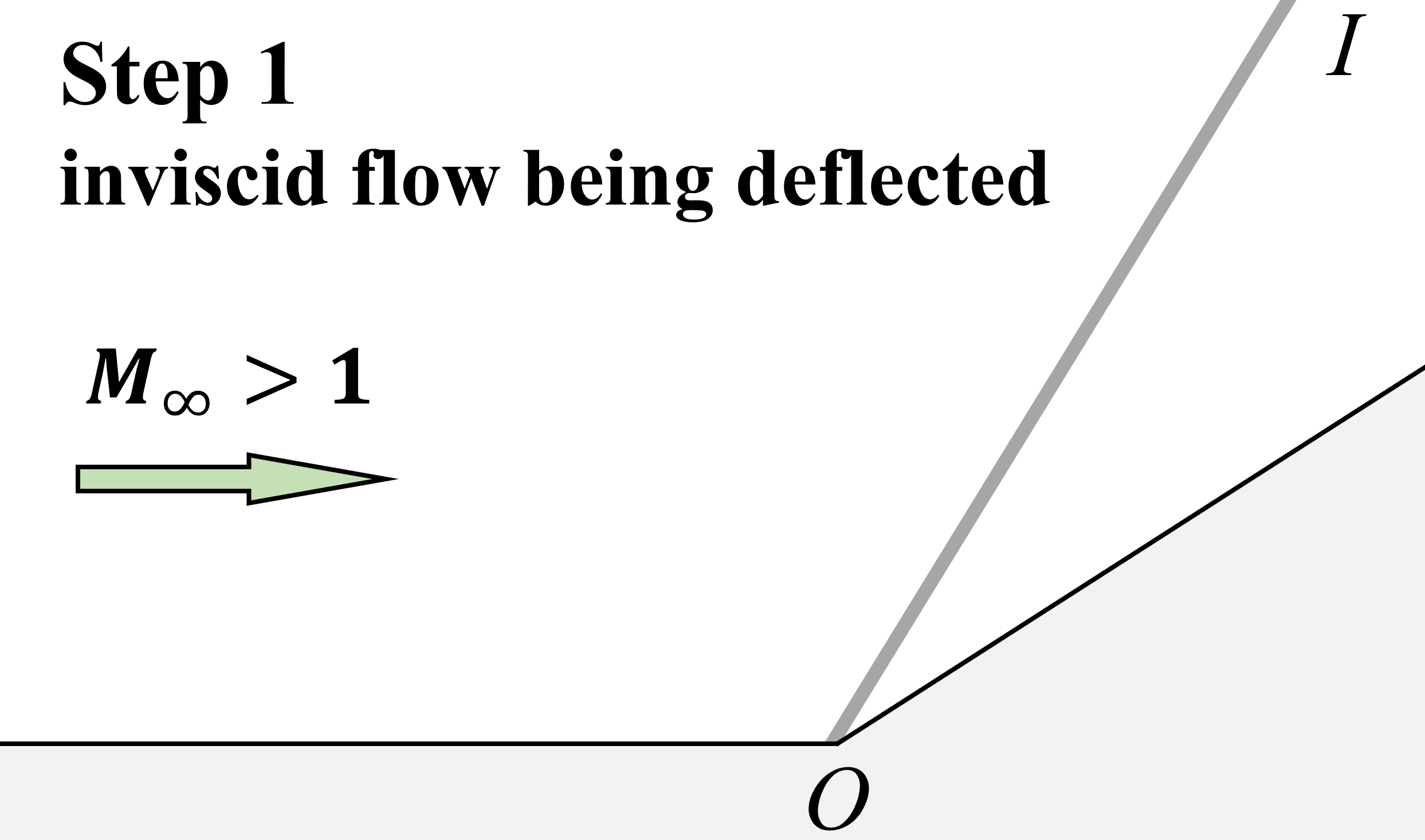}}
	\subfigure[\label{subfig:Process2_Step2}]{\includegraphics[width = 0.32\columnwidth]{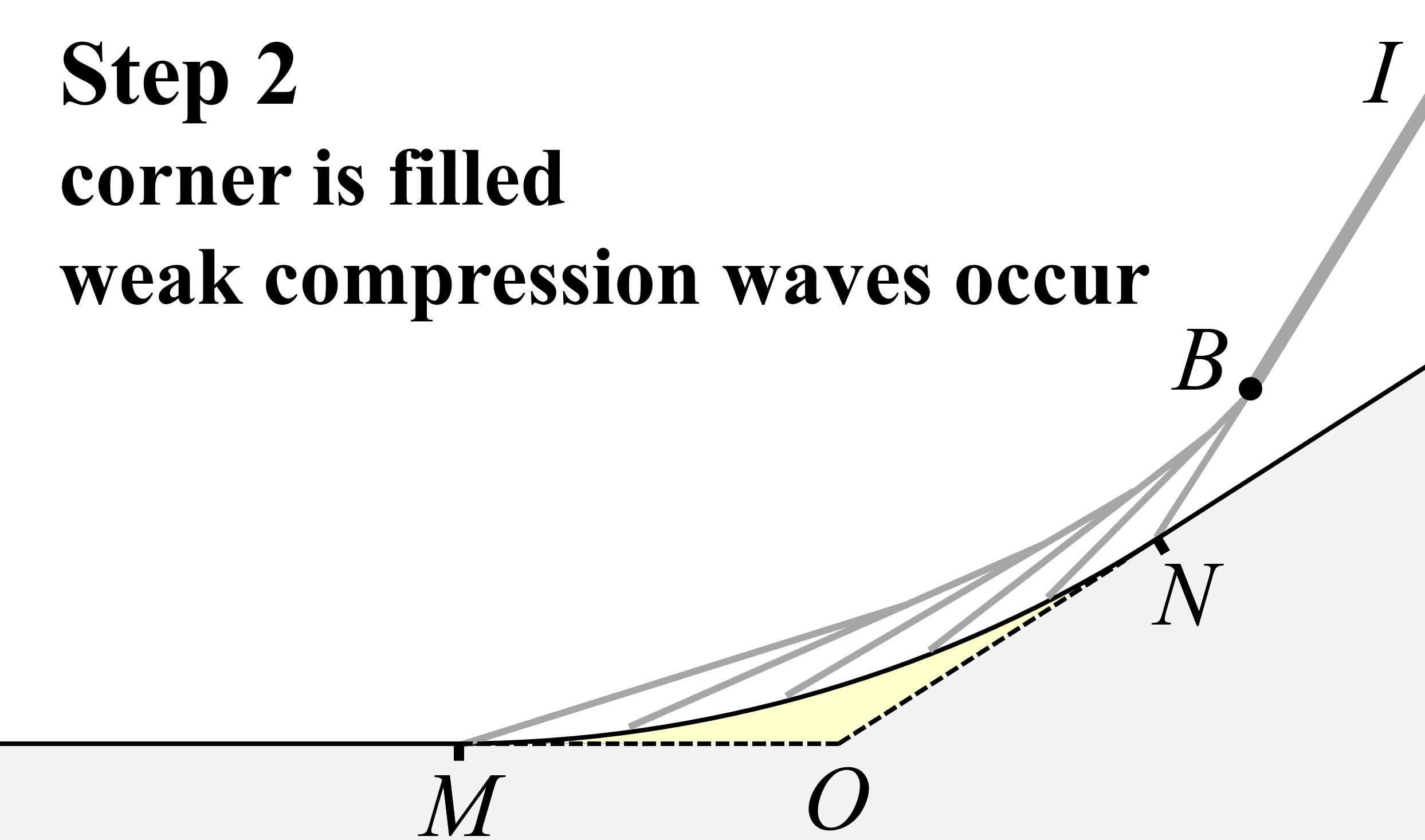}}
	\subfigure[\label{subfig:Process2_Step3}]{\includegraphics[width = 0.32\columnwidth]{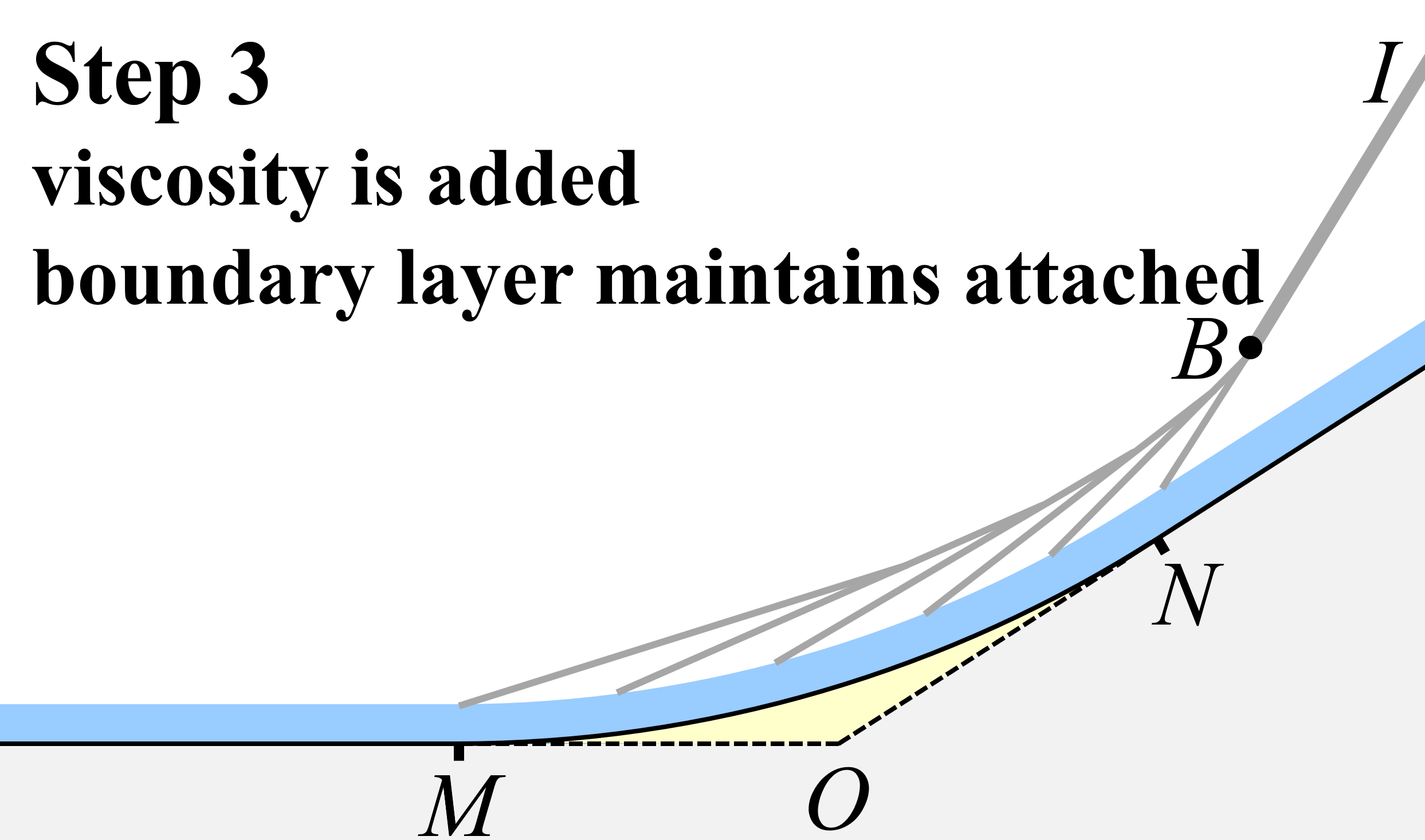}}
	\caption{Process 2 to construct an attached CCR flow. (a)Step 1;(b)step 2;(c)step 3.}
\end{figure}

\begin{itemize}
	\item {\bf Step 1: deflect the inviscid flow.} This step is the same as {\bf Step 1} in process 1;
	\item {\bf Step 2: fill the corner as an arc.} This step is similar to {\bf Step 3} in process 1, and the only difference is that the corner's filling comes before the fluid viscosity's adding. Due to the curved wall $\overline{MN}$, SW $OI$ emitting from point $O$ is weakened to a series of compression waves (CWs) spreading on $\overline{MN}$, as shown in Fig. \ref{subfig:Process2_Step2};
	\item {\bf Step 3: add the viscosity.} This step is similar to {\bf Step 2} in process 1. The viscosity is suddenly added to the fluid after the inviscid flow stably passing over the CCR. Undergoing the weaker adverse pressure gradients of CWs on $\overline{MN}$, the gradually formed BL is very likely to stabilize at attachment state, as shown in Fig. \ref{subfig:Process2_Step3}.
\end{itemize}
\par Compare processes 1 and 2. Operationally, both SA and SS CCR flows could be obtained with three steps by only exchanging the order of the last two steps. Essentially, the different presentations of the ultimate stable states originate from the emergence history of flow structures, including BL, SW and CW. In Sec. \ref{sec:Numerical expriment}, we will use DNSs to check the authenticity of the thought experiment.
\label{subsec:Conjecture}

\section {Numerical expriment}
\label{sec:Numerical expriment}
\par 3D DNSs are performed to realize the above thought experiment in Sec. \ref{subsec:Numerical expriment} as much as possible. The DNSs are conducted with OpenCFD, whose capability has been well examined \citep{hu2017beta,xu2021effect}. Details of the numerical methods, the verification, and the validation are presented in Sec. \ref{subsec:verification}.
\subsection {Reproduction of the thought experiment}
\label{subsec:Numerical expriment}
The flow conditions of the DNSs are mainly based on the recent shock-tunnel experiments \citep{roghelia2017experimental,Roghelia2017ExperimentalIO,chuvakhov2017effect}. The test flow is an SS DCR with flat plate length $L = 100$mm and $\alpha = 15^{\circ}$. The inflow and wall parameters of the test gas (air) include the Mach number $M_{\infty} =7.7$, the Reynolds number $Re_{\infty}(\text{m}) = 4.2 \times 10^{6}$, the velocity $u_{\infty} = 1745 \text{m/s}$, the density $\rho_{\infty} = 0.021 \text{kg m}^{-3}$, the pressure $p_{\infty} = 760 \text{Pa}$, the temperature $T_{w} = 125 \text{K}$, the wall temperature $T_{\infty} = 293 \text{K}$, the Prandtl number $Pr = 0.7$, and the specific heat ratio $\gamma = 1.4$. The viscosity is calculated with the Sutherland law.
\begin{figure}
	\centering
	\subfigure[\label{subfig:Numerical_Process1-step2}]{\includegraphics[width = 0.49\columnwidth]{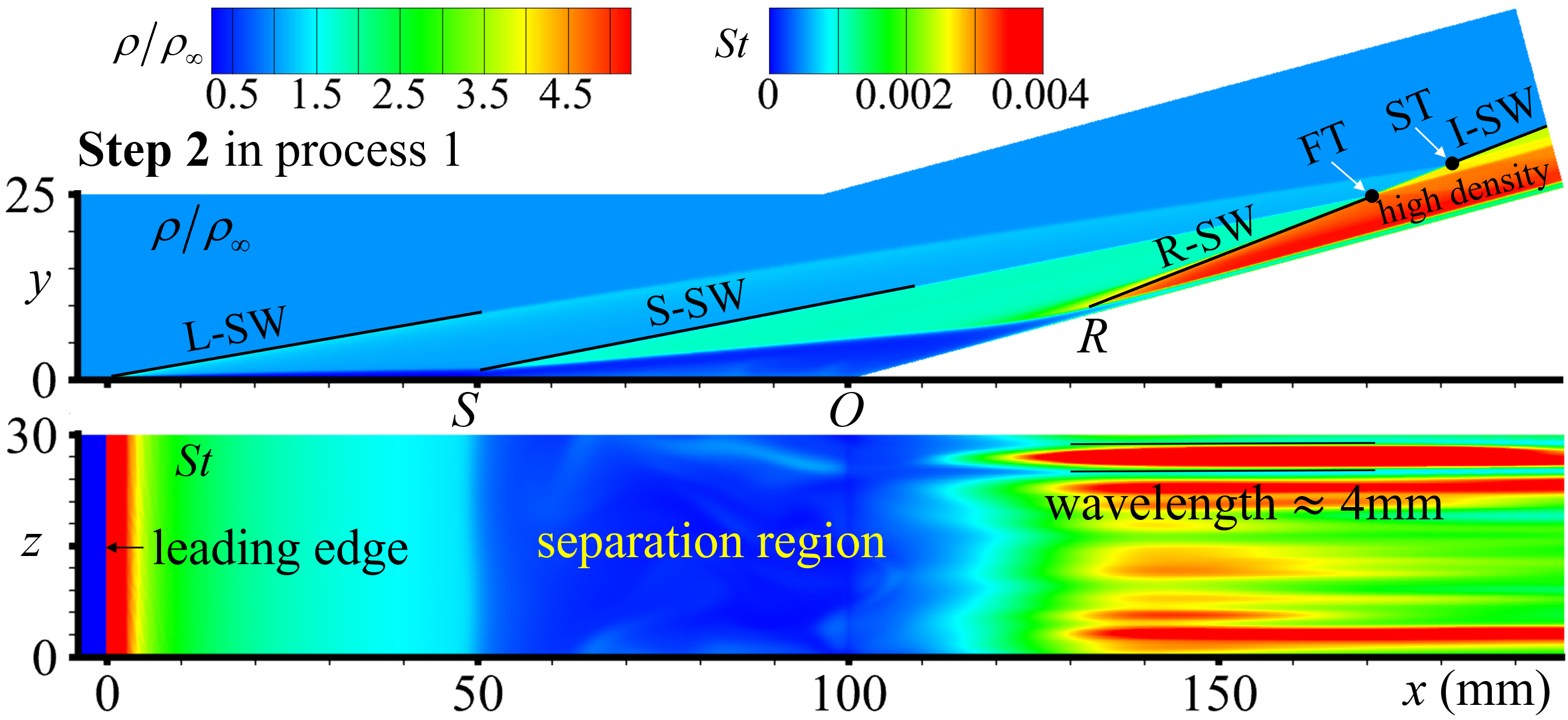}} 
	\subfigure[\label{subfig:Numerical_Process1-step3}]{\includegraphics[width = 0.49\columnwidth]{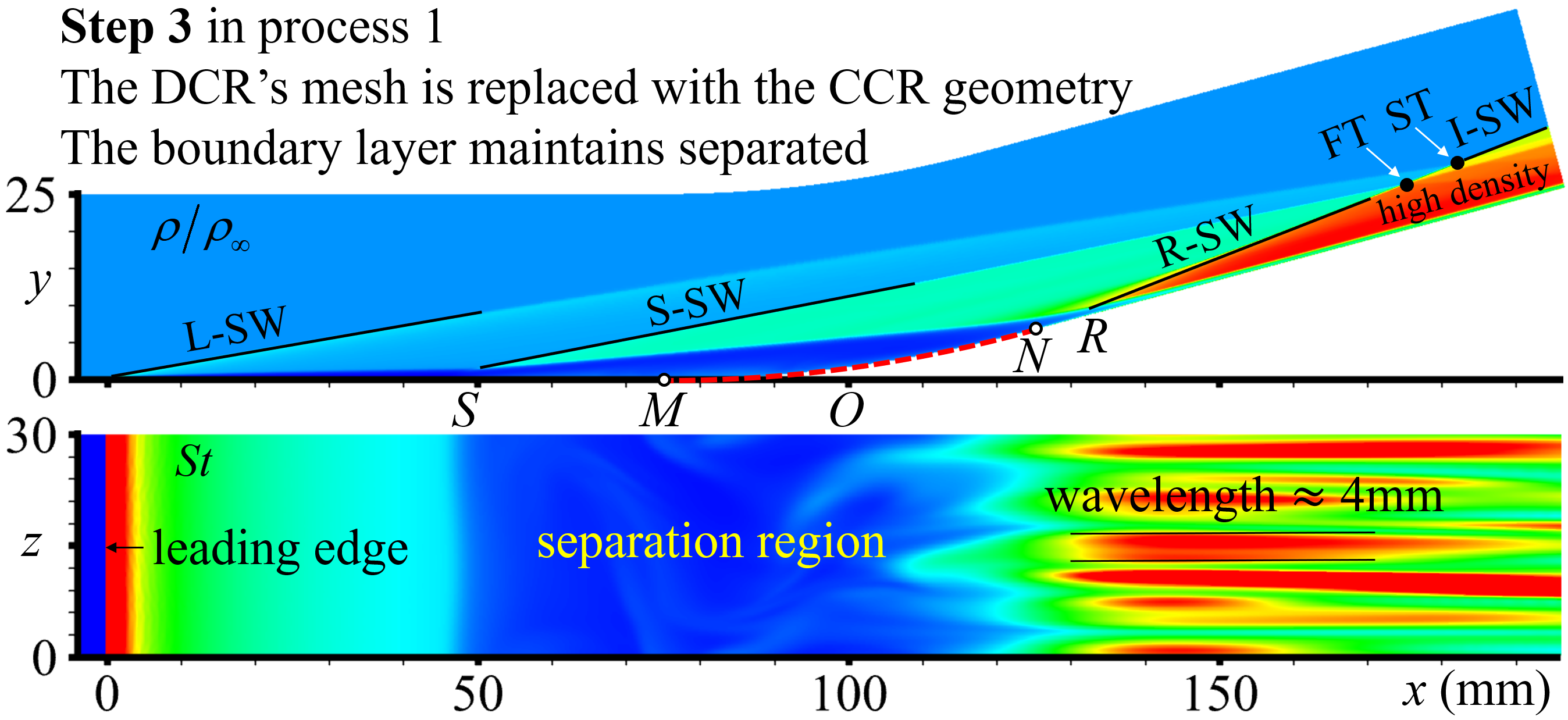}} 
	\subfigure[\label{subfig:Numerical_Process2-step2}]{\includegraphics[width = 0.49\columnwidth]{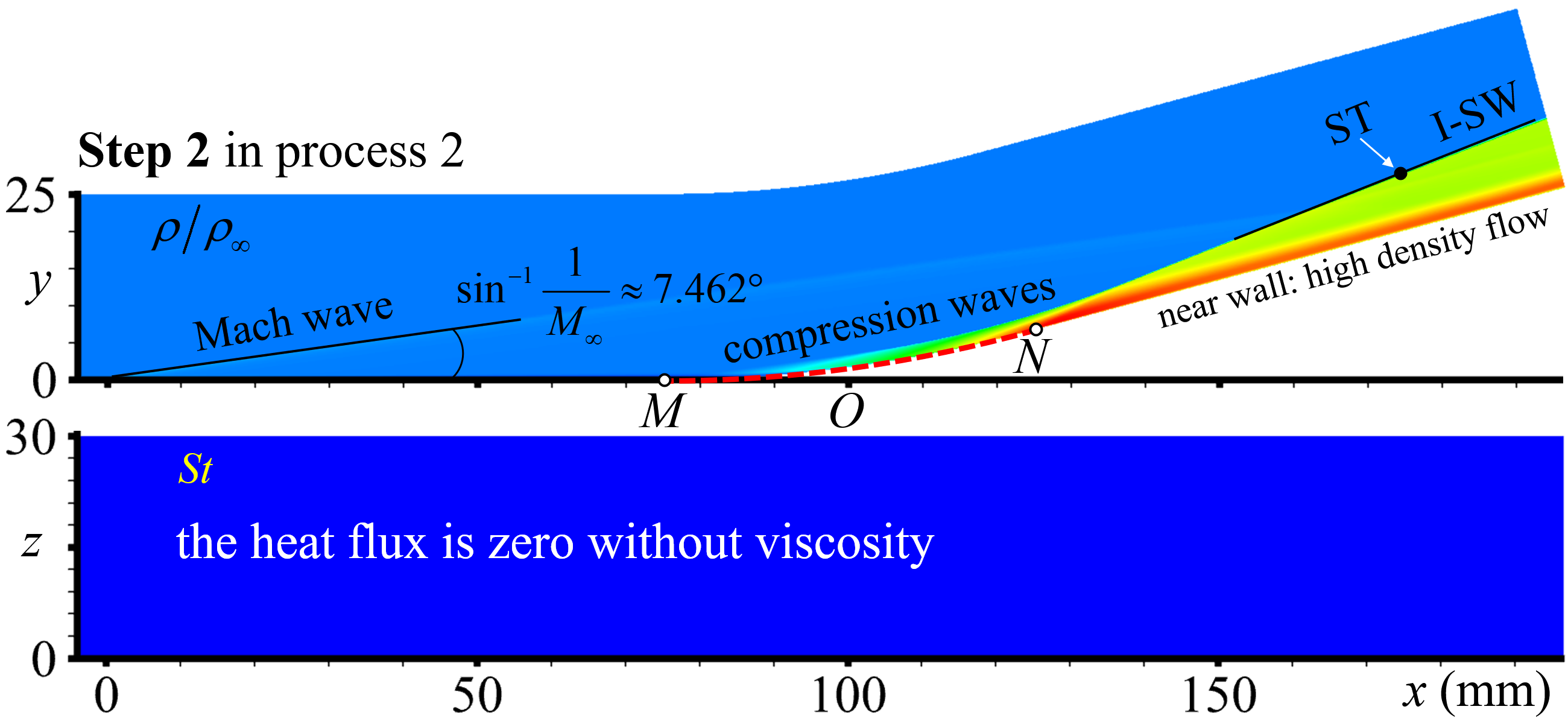}} 
	\subfigure[\label{subfig:Numerical_Process2-step3}]{\includegraphics[width = 0.49\columnwidth]{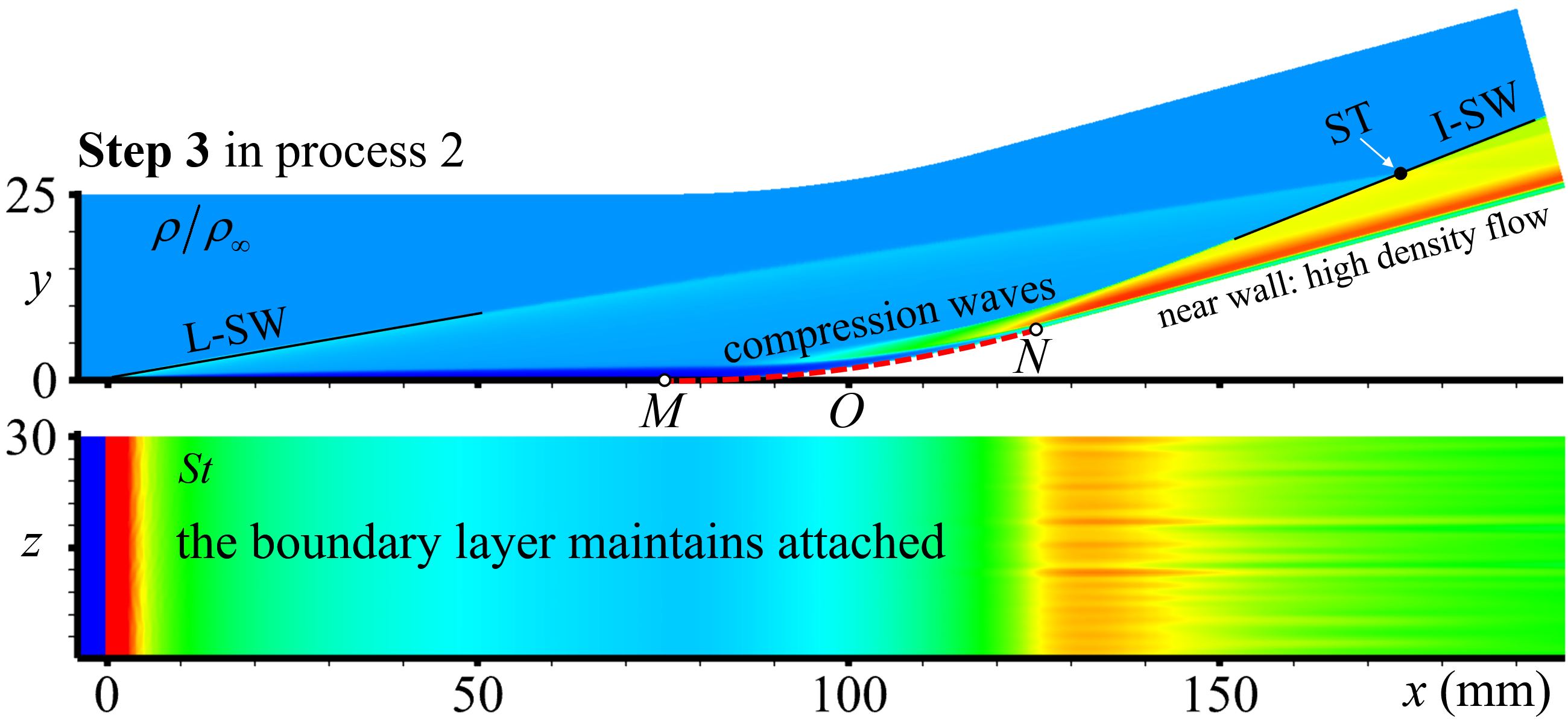}} 
	\caption{Numerical experiments to reproduce (a) Step 2 and (b) Step 3 in process 1; (c) Step 2  and (d) Step 3 in process 2.}\label{fig:DNS}
\end{figure}
\par First, we reproduce process 1. Steps 2 in process 1 can be reproduced using the original SS DCR, whose simulation results (colored by density in $x-y$ plane and normalized wall heat flux, the Stanton number $St$, in $x-z$ plane) are shown in Fig. \ref{subfig:Numerical_Process1-step2}. The SWs includes the leading edge SW (L-SW), the separation SW (S-SW), the reattachment SW (R-SW), and the inviscid SW (I-SW). The first triple (FT) point is the intersection of S-SW and R-SW, and the second triple (ST) point is the intersection of L-SW and I-SW. The wavelengths of the spanwise streaks are about $4 \sim 6$mm, which are consistent with the previous experimental observations \citep{roghelia2017experimental,Roghelia2017ExperimentalIO,chuvakhov2017effect} and numerical results \citep{cao2021unsteady}. Considering the corner is filled into an arc with an “imaginary machine” in step 3, we directly replace the DCR’s mesh with the CCR geometry ($MO = ON = 25$mm), numerically. This operation actually leads to a more intense perturbance than the “imaginary machine” and less conducive to obtaining an SS CCR flow. Even so, as shown in Fig. \ref{subfig:Numerical_Process1-step3}, the obtained CCR flow finally stabilizes in the separation state, whose simulation lasted for nearly 20ms, and the animation can be seen in Movie 1.
\par Then we reproduce process 2. Numerically, impermeable and free-slip wall conditions are set to realize step 2 in process 2, and the obtained flow field are shown in Fig. \ref{subfig:Numerical_Process2-step2}. A series of CWs distribute on the curved wall $\overline{MN}$ and induce the near-wall high density flow in the downstream. Since step 3 in process 2 is adding viscosity into the fluid, we numerically change the BC from free-slip wall to non-slip wall. As shown in Fig. \ref{subfig:Numerical_Process2-step3}, the obtained CCR flow finally stabilizes in the attachment state for more than 5ms (see also Movie 2).
\par Compare Fig. \ref{subfig:Numerical_Process1-step3} and Fig. \ref{subfig:Numerical_Process2-step3}, the distinct-different stable states are under the same BCs, verifying the existence of the bistable states in CCR flows. Also note that the 3D effects and local unsteadiness emerge in the downstream, manifesting as the 3D structures in the separation bubble and spanwise streaks on the ramp. This implies the bistatble states can resist disturbances of a certain intensity. Till now, we have verified the existence of the bistable states. However, the rationality for bistable states’ existence remains unexplained. In Sec. \ref{sec:Demonstration}, the bistability's inevitability will be demonstrated theoretically.
\subsection{Verification and validation}
\label{subsec:verification}
\par As the computational domain shown in Fig. \ref{fig:DNS}, the streamwise region is $-2 \text{mm} \le x \le 196.6 \text{mm}$ (the leading edge is at $0$ mm), the wall-normal height is $25$ mm, and the spanwise width is $30$ mm. Two mesh resolutions, $1020 \times 250 \times 300$ (M1) and $1420 \times 250 \times 300$ (M2), are considered for time and grid convergence studies, whose initial flow fields are both obtained from the two-dimensional simulations with $x_{S}/L \approx 0.55$ and $x_{R}/L \approx 1.28$. Both M1 and M2 cluster the grids near the wall with the first wall-normal grid height being fixed to $\Delta y_{\text{w}} = 0.008$ mm, yielding the non-dimensional height $\Delta y_{\text{w}}^{+} \approx 0.3$ mm at $x/L \approx 0.5$. The spanwise grid spaces of the two meshes are both uniform as $\Delta z = 0.1$ mm, which is smaller than $0.125$ mm of \citet{cao2021unsteady}. The streamwise grid spaces are uniform as $\Delta x_{\xi} = 0.2$ mm in M1. To investigate the pressure gradient effects, the spaces of $\Delta x_{\xi}$ are both clustered near $x_{S}/L \approx 0.5$ and $x_{R}/L \approx 1.3$ with $200$ points in M2. In terms of the numerical methods, the time integration is performed by the third-order TVD-type Runge–Kutta method; the inviscid fluxes are discreted by the fifth-order WENO method \citep{jiang1996efficient}; the viscid fluxes are discreted with the sixth-order central difference scheme.
\par
\begin{figure}
	\centering
	\subfigure[\label{subfig:TimeConvergence}]{\includegraphics[width = 0.45\columnwidth]{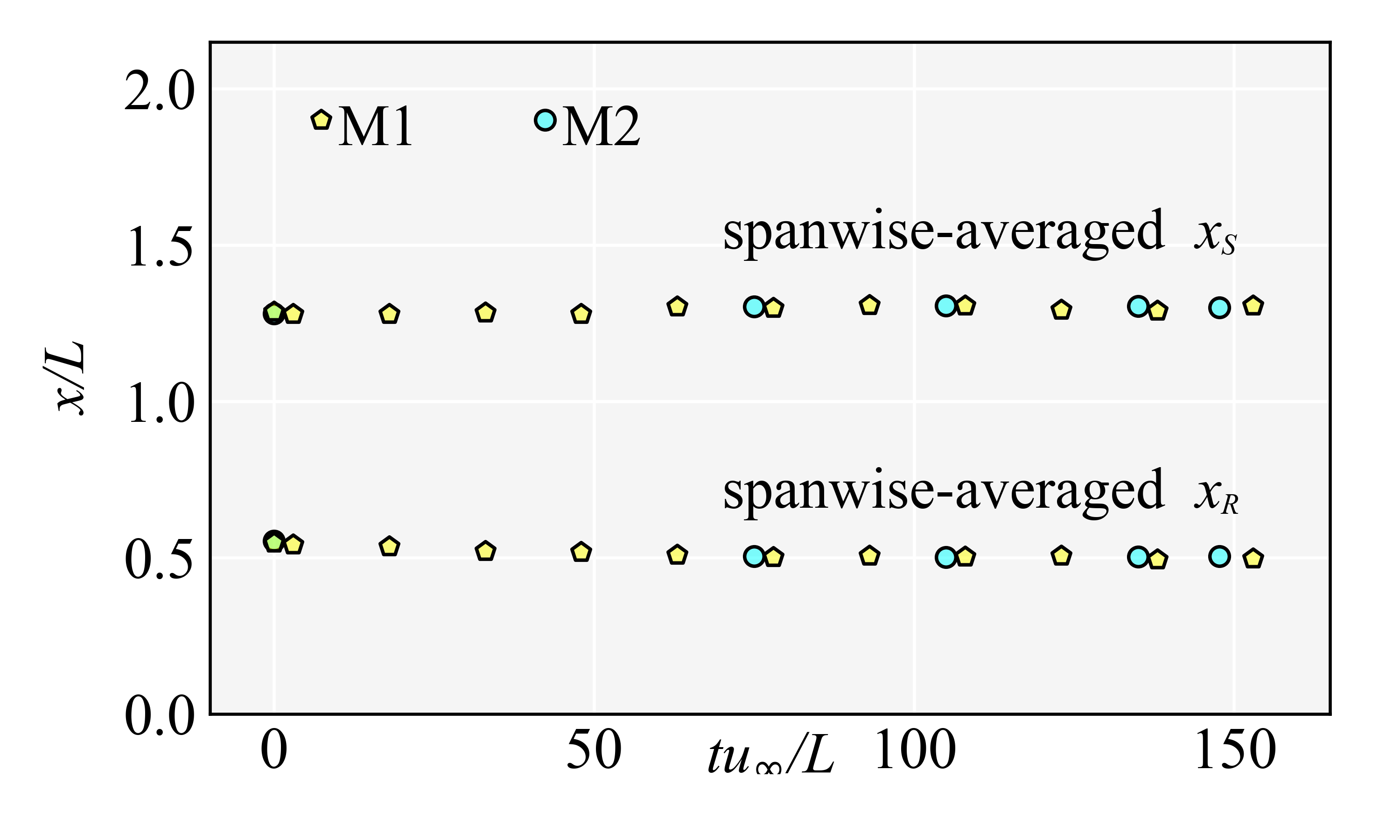}}
	\subfigure[\label{subfig:Validation_Cp_St_Cf}]{\includegraphics[width = 0.45\columnwidth]{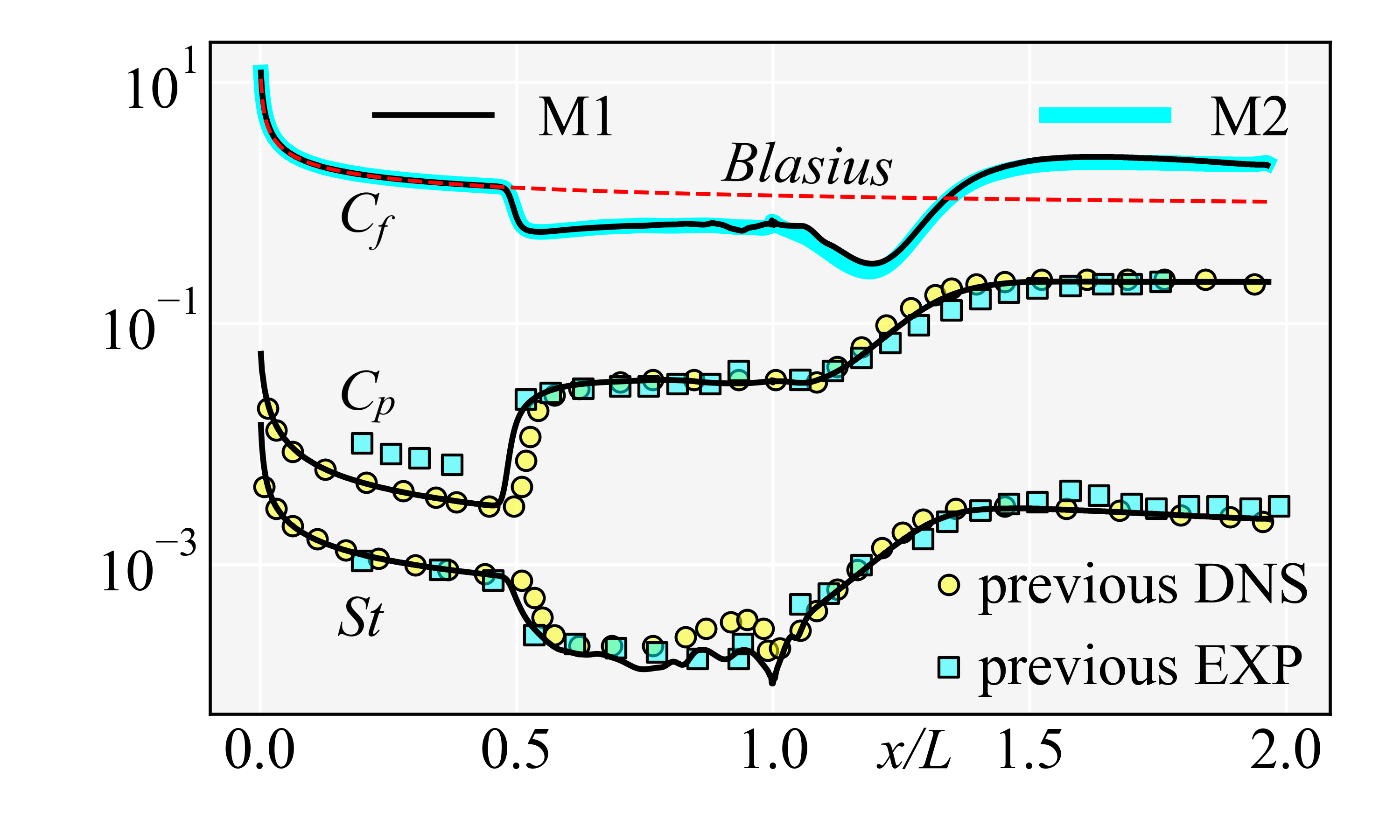}}
	\vspace{-2.5mm}
	\caption{(a) Variations of $x_{S}$ and $x_{R}$ with time; (b) distributions of $C_p$, $C_p$, and $St$.}
\end{figure}

 As shown in Fig. \ref{subfig:TimeConvergence}, the spanwise-averaged separation and reattachment points, $x_{S}$ and $x_{R}$, have only slight oscillations within 8.3 ms (non-dimensional time $tu_{\infty}/L \approx 145$, see also Movie 3), verifying the time-convergence. Fig. \ref{subfig:Validation_Cp_St_Cf} shows the time-convergent distributions of the skin friction coefficient $C_{f}$, the surface pressure coefficient $C_{p}$, and the wall Stanton number $St$, which are defined as
\begin{equation}
	C_f=\frac{\tau_{w}}{\frac{1}{2}\rho_{\infty} u_{\infty}^2}, \quad C_p=\frac{p_{w} - p_{\infty}}{\frac{1}{2}\rho_{\infty} u_{\infty}^2}, \quad St = \frac{q_w}{\rho_{\infty} u_{\infty} c_{p}(T_{aw} - T_{w})}
	\label{Helm}
\end{equation}
where $\tau_{w}$, $p_{w}$ and $q_{w}$ are the friction, pressure and heat flux of the surface, respectively. $c_p$ and $T_{aw}$ are the specific heat capacity and the adiabatic wall temperature, respectively. The small discrepancy between the $C_f$ distributions of M1 and M2 verifies the simulations' grid-convergence (M1), both of which agree well with the Blasius theoretical solution \citep{white2006viscous} in the flat plate region. The present $C_p$ and $St$ distributions are also compared with the experimental \citep{roghelia2017experimental,chuvakhov2017effect} and previous DNS \citep{cao2021unsteady} results, and the excellent agreements validate the present simulations.
\section{Theoretical demonstration}
\label{sec:Demonstration}
\par In this section, we demonstrate DCR flows’ monostability and CCR flows’ bistability. 
\par First, the CR flows' structural features are simplified with the following three ansatzes:
\par {\bf (i) The quasi two-dimensional separation is in the closed form.} This implies that the points $S$ and $R$ appear and disappear together \citep{babinsky2011shock};
\par {\bf (ii) The path $SR$ of the free shear layer (FSL) is approximately simplified as a straight line.} This is based on a lot of observations \citep{helm2021characterization} and the principle of least action \citep{hu2020prediction}, i.e., the FSL should be kept as straight as possible to avoid inducing non-physical SWs that could increase entropy production or dissipation;
\par {\bf (iii) The BL and the FSL are simplified as one-dimensional lines sliced by points $S$ and $R$ when observed from the integral scale $\textit{O}(L)$.} For high $Re_{\infty}$, the BL's thickness $\delta_{x}$ satisfies $\delta_{x} = 5.0Re_{\infty}^{-1/2} x ^{1/2} \ll L$ \citep{white2006viscous}. According to the triple-deck theory \citep{stewartson1969self,neiland1969theory}, the characteristic scale $\delta_{S}$ of the separation process, i.e., the distance from the interaction origin to the separation point \citep{babinsky2011shock}, is $\textit{O}(Re_{\infty}^{-3/8} L^{5/8})$, and the reattachment process's scale $\delta_{R}$ is $\textit{O}(Re_{\infty}^{-1/2} L^{1/2})$. Thus, $\delta_{S}$ and $\delta_{R}$ can be respectively reduced to points $S$ and $R$ relative to $L$.
\par Second, two characteristic pressure rises (PRs) are clarified for an ABL:
\par {\bf (i) Imposed PR $\mathcal{A}$:} the PR across an SW or a series of CWs;
\par {\bf (ii) Resistant PR $\mathcal{I}$:} the strongest PR the stable ABL can resist. This is actually the onset PR of incipient separation \citep{babinsky2011shock}, and separation will occur if $\mathcal{I} < \mathcal{A}$.
\par Thus, $\mathcal{I}$ is an intrinsic property of ABL, and $\mathcal{A}$ is ABL's extrinsic pressure condition. According to the free-interation theory \citep{chapman1958investigation,babinsky2011shock}, the resistant PR $\mathcal{I}_{x}$ at $x$ is of the following form:
\begin{equation}
	\label{eq:Resistant PR}
	\mathcal{I}_{x} = \frac{p_{\text{sep}}-p_{\infty}}{p_{\infty}} = \kappa \frac{M_{\infty}^2}{(M_{\infty}^{2}-1)^{1/4}} C_{f,x}^{1/2}
\end{equation}	
where $p_{\text{sep}}-p_{\infty}$ would be the PR if the BL was separated at $x$, and $\kappa $ is suggested to be a constant ($\kappa \approx 1.5$) for laminar flat-plate BL. Thus, $\mathcal{I}_{x}$ is a monotone increasing function of wall friction $C_{f,x}$. On the other hand, $\mathcal{A}_{x}$ can be obtained by the classical Rankine-Hugoniot and Prandtl-Meyer relations, which are both monotone increasing functions of the defected angle $\phi$ of the flow. Then we have
\begin{equation}
\label{eq:inequation}
\forall {x_1},{x_2} \in ({0,x_{O}}]:\left\{ \begin{array}{l}
	{C_{f,x_1}} > {C_{f,x_2}} \Rightarrow \mathcal{I}_{x_1} > \mathcal{I}_{x_2}\\
	\phi_{x_1}  > \phi_{x_2}  \Rightarrow \mathcal{A}_{x_1}  > \mathcal{A}_{x_2}.
\end{array} \right.
\end{equation}
According the definitions of $\mathcal{I}$ and $\mathcal{A}$, for a given ABL, we can obtain
\begin{equation}
	\label{eq:I-A_relation}
	\left\{\begin{array}{lll}
		\forall x > 0: \mathcal{I}_x \geq \mathcal{A}_x &\Leftrightarrow \text{the ABL will maintain attached;} \\
		\exists x > 0: \mathcal{I}_x<\mathcal{A}_x &\Leftrightarrow \text{the ABL will separate.} 
	\end{array}\right.
\end{equation}
According to ansatze (iii), for a BL stably separating at $x_{S}$ with negligible $\delta_{S}$, the separation PR can be accomplished only in one point $x_{S}$ when observed from the integral scale. Thus, $\mathcal{I}$ should be balanced with $\mathcal{A}$ at $x_{S}$:
\begin{equation}
	\label{eq:separation}
	\mathcal{I}_{S} = \mathcal{A}_{S}.
\end{equation}
\par Next, we will demonstrate DCR's monostability and CCR's bistability using the method of virtual separation disturbance (VSD). The core thought is to impose a VSD satisfying the geometric constraints into the SA states, and then check if it can be stimulated. The SA state with a stimulable VSD has a potential SS state.
\subsection{The proof of DCR's monostability}
\par Proving DCR's monostability is equivalent to proving: (i) for an SA state, there is no potential SS state; {\bf or} (ii) for an SS state, there is no potential SA state. Here we prove case (i).
\par
\begin{figure}
	\centering
	\subfigure[\label{subfig:SA_state_Process1}]{\includegraphics[width = 0.32\columnwidth]{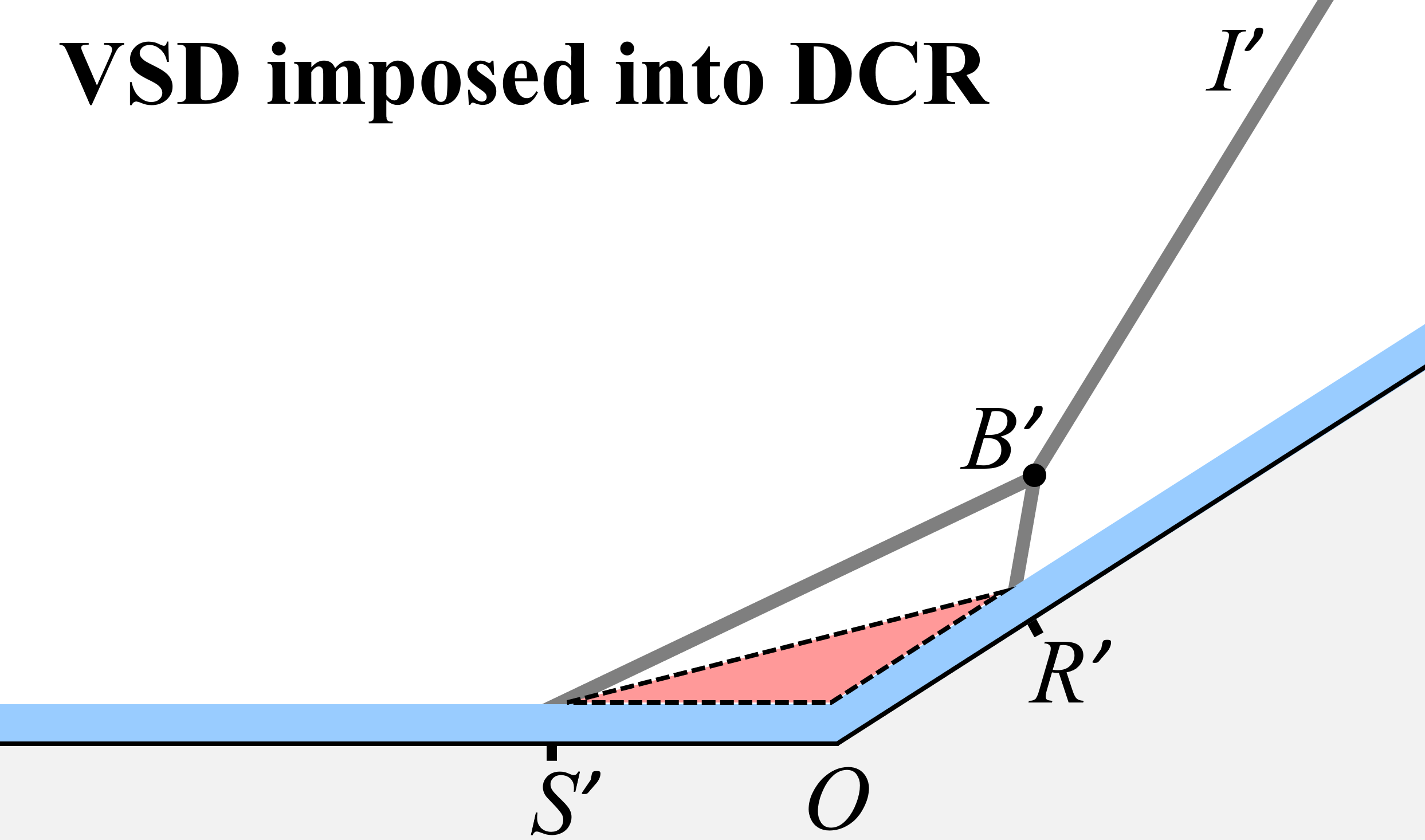}}
	\subfigure[\label{subfig:VSD_CCR_SA}]{\includegraphics[width = 0.32\columnwidth]{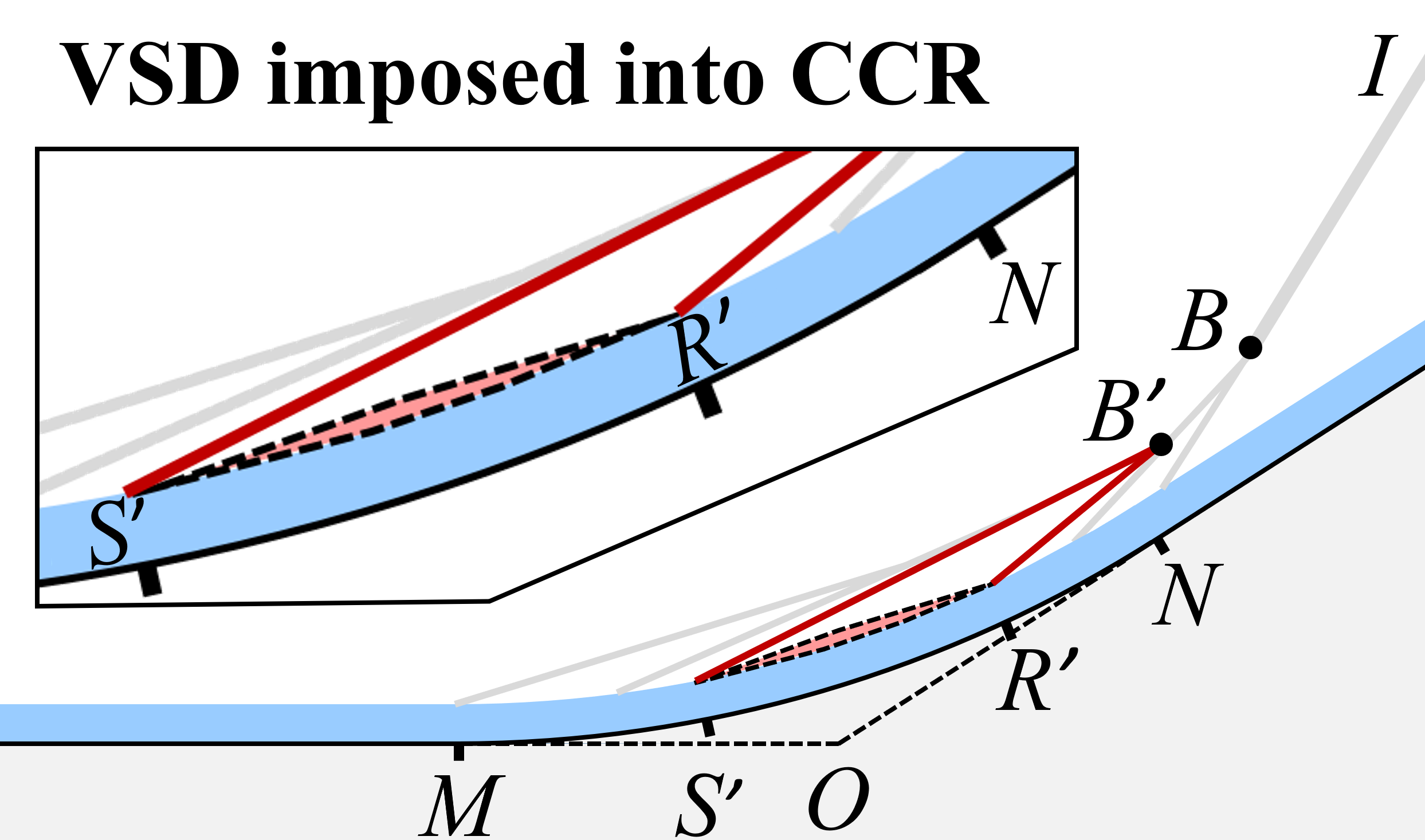}}
	\subfigure[\label{subfig:DCR_CCR_Cf}]{\includegraphics[width = 0.32\columnwidth]{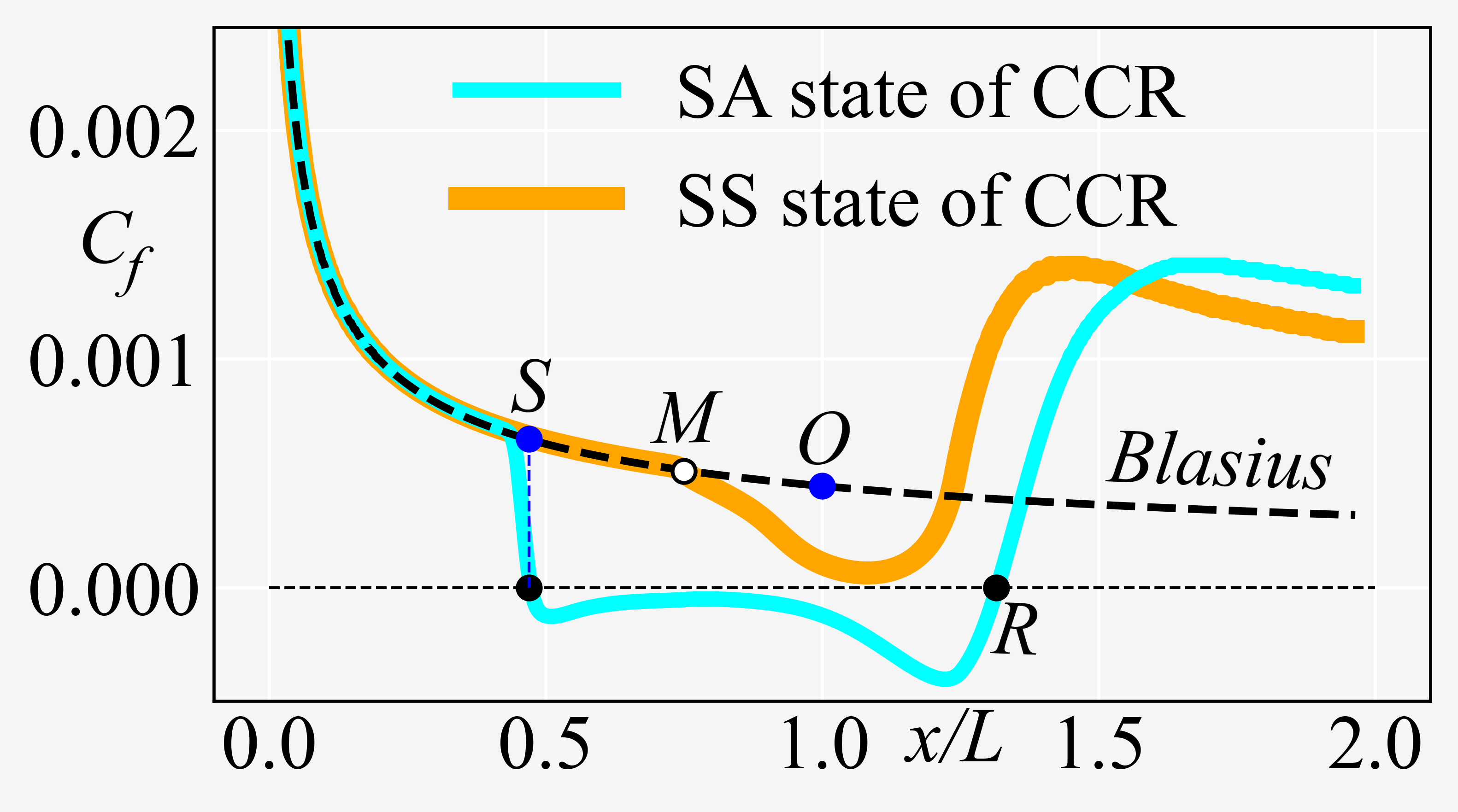}}
	\caption{SW patterns of (a) DCR and (b) CCR with VSD; (c) $C_{f}$ of SS and SA states.}
\end{figure}

 For an SA DCR flow, according to relations (\ref{eq:I-A_relation}), there must be 
\begin{equation}
\label{eq:DCR_O}
\mathcal{I}_{O}^{\text{D}} \geq \mathcal{A}_{O}^{\text{D}}.
\end{equation}
where superscript `D' denotes DCR geometry. As shown in Fig. \ref{subfig:SA_state_Process1}, a VSD $\triangle{S'O'R'}$ is imposed to the SA DCR flow, inducing the virtual SWs $S'B'$ and $R'B'$ at $x_{S'}$ and $x_{R'}$, respectively. Unlike a real separation bubble covered by the FSL, the VSD is designed as an adaptive isobaric dead-air region upon the ABL, which only enlarges the outer inviscid flow's deflection and changes the shock configuration but without changing the ABL's properties. According to ansatze (ii), $S'R'$ is a straight line representing the FSL, then the VSD can be depicted with $x_{S'}$ and $\phi_{S'}$ (the angle between $S'R'$ and the $x-$axis), which satisfies $\phi_{S'} = \angle R'S'O < \alpha =  \phi_{O}$. For all possible VSDs, denote the collection of $\mathcal{A}_{S'}$ as complete set $\mathsfbi{A}$, then we have
\begin{equation}
	\label{eq:DCR_VSD}
	\mathcal{A}_{S'}^{\text{D}} \in \mathsfbi{A}^{\text{D}} := \left\{(x_{S'}, \phi_{S'}) \mid x_{S'}\in (0,x_{O}), \phi_{S'} \in [0,\alpha) \right\}.
\end{equation}
According inequations (\ref{eq:inequation}) and the theoretical solution $C_{f,x} \propto Re_{x}^{-1/2} \propto x^{-1/2}$ of the laminar flat-plate BL \citep{schlichting2003boundary}, we can obtain
\begin{equation}
	\label{eq:DCR_Phi_Cf}
	\forall \mathcal{A}_{S'}^{\text{D}} \in \mathsfbi{A}^{\text{D}}:\left\{\begin{array}{lll}
		C_{f,S'}^{\text{D}} > C_{f,O}^{\text{D}} &\Rightarrow \mathcal{I}_{S'}^{\text{D}} > \mathcal{I}_{O}^{\text{D}} \\
		\phi_{S'} < \phi_{O} &\Rightarrow \mathcal{A}_{S'}^{\text{D}} < \mathcal{A}_{O}^{\text{D}}.
	\end{array}\right.
\end{equation}
Combine inequations (\ref{eq:DCR_O}) and (\ref{eq:DCR_Phi_Cf}), we can finally obtain
\begin{equation}
	\label{eq:DCR_S'}
	\forall \mathcal{A}_{S'}^{\text{D}} \in \mathsfbi{A}^{\text{D}}: \mathcal{I}_{S'}^{\text{D}} > \mathcal{A}_{S'}^{\text{D}}.
\end{equation}
Inequation (\ref{eq:DCR_S'}) means that for an SA DCR, any VSD will decay due to $\mathcal{I}_{S'}^{\text{D}} > \mathcal{A}_{S'}^{\text{D}}$, and the disturbed DCR flow must return to the SA state. Then case (i) has been proved, and we can get the conclusion that there is only one stable state in DCR for any given BC, which is consistent with empirical observations.

\subsection{The proof of CCR's bistability}
\par Proving CCR's bistability is equivalent to proving: (i) for an SA state, there is a potential SS state; {\bf or} (ii) for an SS state, there is the potential SA state. Here we prove case (i).
\par For an SA CCR flow, according to the relations (\ref{eq:I-A_relation}), there must be 
\begin{equation}
	\label{eq:CCR_SA}
	\forall x \in (0,x_{N}):\mathcal{I}_{x}^{\text{C}} \geq \mathcal{A}_{x}^{\text{C}}
\end{equation}
where superscript `C' denotes CCR geometry. A VSD $\triangle{S'\overline{R'S'}}$ shown in Fig. \ref{subfig:VSD_CCR_SA} is imposed into the SA CCR flow, the space occupied by which is inside space $\overline{MN}B$. Then we have
\begin{equation}
	\label{eq:CCR_VSD}
	\mathcal{A}_{S'}^{\text{C}} \in \mathsfbi{A}^{\text{C}} := \left\{(x_{S'}, \phi_{S'}) \mid x_{S'}\in (0,x_{N}), \phi_{S'} \in [0,\alpha) \right\}
\end{equation}
Similar to the SA DCR, the VSD in CCR also induces a stronger deflection $\phi_{S'}$ than that of its original SA state (denoted by superscript `C,SA'), i.e., $\phi_{S'} > \phi_{S'}^{\text{C,SA}}$, but does not change the BL's $\mathcal{I}$. Then we have 
\begin{equation}
\label{eq:I-A_CCR}
\forall \mathcal{A}_{S'}^{\text{C}} \in \mathsfbi{A}^{\text{C}}:\left\{\begin{array}{lll}
	\mathcal{I}_{S'}^{\text{C}} = \mathcal{I}_{S'}^{\text{C,SA}} \geq \mathcal{A}_{S'}^{\text{C,SA}} & \Leftrightarrow \mathcal{I}_{S'}^{\text{C,SA}} - \mathcal{A}_{S'}^{\text{C,SA}} = \varepsilon \geq 0 \\
	\mathcal{A}_{S'}^{\text{C}} > \mathcal{A}_{S'}^{\text{C,SA}} & \Leftrightarrow \mathcal{A}_{S'}^{\text{C}} - \mathcal{A}_{S'}^{\text{C,SA}} = \varepsilon' > 0
\end{array}\right.
\end{equation}
where $\varepsilon$ measures the residual PR of the original SA BL, and $\varepsilon'$ is the extra PR induced by the VSD. Then we finally obtain 
\begin{equation}
	\label{eq:uncertain}
	\mathcal{A}_{S'}^{\text{C}} - \mathcal{I}_{S'}^{\text{C}} = \varepsilon' - \varepsilon
\end{equation}
Due to the arbitrariness of $\varepsilon'$, the magnitude relationship between $\mathcal{I}_{S'}^{\text{C}}$ and $\mathcal{A}_{S'}^{\text{C}}$ can not be directly obtained, which is different from DCR. Further, there are three possible subcases:
\begin{equation} 
	\label{eq:Bistability_CCR}
	\left\{\begin{aligned}
		\text{(i-i)} \quad & \forall \mathcal{A}_{S'}^{\text{C}} \in \mathsfbi{A}^{\text{C}}: \mathcal{A}_{S'}^{\text{C}} \leq \mathcal{I}_{S'}^{\text{C}} \Leftrightarrow \varepsilon' \leq \varepsilon \\
		\text{(i-ii)} \quad & \forall \mathcal{A}_{S'}^{\text{C}} \in \mathsfbi{A}^{\text{C}}: \mathcal{A}_{S'}^{\text{C}} > \mathcal{I}_{S'}^{\text{C}} \Leftrightarrow \varepsilon' > \varepsilon \\
		\text{(i-iii)} \quad & \exists \mathsfbi{A_{ps}} \subsetneqq \mathsfbi{A}^{\text{C}}: \left\{\begin{array}{l}
			\forall \mathcal{A}_{S'}^{\text{C}} \in \mathsfbi{A_{ps}}: \mathcal{A}_{S'}^{\text{C}} \leq \mathcal{I}_{S'}^{\text{C}} \Leftrightarrow \varepsilon' \leq \varepsilon \\
			\forall \mathcal{A}_{S'}^{\text{C}} \in \mathsfbi{A_{ps}^{C}}: \mathcal{A}_{S'}^{\text{C}} > \mathcal{I}_{S'}^{\text{C}} \Leftrightarrow \varepsilon' > \varepsilon
		\end{array}\right.
	\end{aligned}\right.
\end{equation}
where $\mathsfbi{A_{ps}}$ is a proper subset as $\mathsfbi{A}^{\text{C}}$, whose complementary set is $\mathsfbi{A_{ps}^{C}}$. Subcase (i-i) is similar to inequation (\ref{eq:CCR_SA}), implying any VSD can not be maintained and will disappear gradually, and no potential SS state exists. Subcase (i-ii) means any VSD can be reinforced, implying the original SA state is a critical state only existing in the ideal condition without disturbance, which can not be observed in reality. In subcase (i-iii), on the one hand, a series of VSDs in $\mathsfbi{A_{ps}^{C}}$ will be maintained and reinforced; on the other hand, the CCR flow also has the ability to resist some disturbances $\mathsfbi{A_{ps}}$, consisting with the stable existence of the original undisturbed SA state. Subcase (i-iii) explains the occurrence of bistable states, pointing out the existence's possibility. Next, for a specific case, we will prove the existence's inevitability. 
\par Consider an SA CCR flow with ramp angle $\alpha$ and curved wall $\overline{MN}$, and its corresponding DCR flow (with the same $\alpha$ and inflow conditions) being at SS state. According to (\ref{eq:I-A_CCR}) and (\ref{eq:separation}), we have
\begin{equation}
	\label{eq:DCR_S_rewritten}
	\mathcal{I}_{M}^{\text{C,SA}} - \mathcal{A}_{M}^{\text{C,SA}} = \varepsilon_{M} \ge 0, \quad \mathcal{I}_{S}^{\text{D}} = \mathcal{A}_{S}^{\text{D}}
\end{equation}
As shown in \ref{subfig:DCR_CCR_Cf}, according to relations (\ref{eq:DCR_Phi_Cf}) and the Blasius theoretical resolution (denoted by superscript `B'), we have
\begin{equation}
	\label{eq:DCR_S_Q}
	C_{f,M}^{\text{C,SA}} = C_{f,M}^{\text{B}} < C_{f,S}^{\text{B}} = C_{f,S}^{\text{D}} \Rightarrow \mathcal{I}_{M}^{\text{C,SA}} < \mathcal{I}_{S}^{\text{D}}.
\end{equation}
Combine (\ref{eq:DCR_S_rewritten}) and (\ref{eq:DCR_S_Q}), we can obtain $\mathcal{A}_{S}^{\text{D}} - \mathcal{A}_{M}^{\text{C,SA}} >\varepsilon_{M}$. Thus, if a VSD with PR $\mathcal{A}_{M}^{\text{C}} = \mathcal{A}_{S}^{\text{D}}$ can be imposed at $x_{M}$, there must be
\begin{equation}
	\label{eq:impose_PR_E}
	\mathcal{A}_{M}^{\text{C}} - \mathcal{A}_{M}^{\text{C,SA}} = \varepsilon'_{M}>\varepsilon_{M}.
\end{equation}
According to subcase (i-iii) in (\ref{eq:Bistability_CCR}), the VSD satisfying (\ref{eq:impose_PR_E}) will be stimulated. For the SS DCR flow, the deflection $\phi_{S}^{\text{D}}$ induced by the separation bubble can be calculated quantitatively \citep{chapman1958investigation,hu2020prediction}, and must be smaller than $\alpha$, i.e., $\phi_{S}^{\text{D}} \in (0,\alpha)$. Let the VSD imposed at $x_{M}$ be with $\phi_{M} = \phi_{S}^{\text{D}} \in [0,\alpha)$, then geometric constraint (\ref{eq:CCR_VSD}) can be satisfied, and the VSD must be stimulated. Till now, the inevitability of the existence of bistable states has been proved, which is consistant with the thought experiment and the numerical results. Subcase (i-iii) is actually the necessary and sufficient conditions for the emergence of bistable states.
\subsection{Discussion}
\par The essential reason for the imparity between DCR's monostability and CCR's bistability is the difference in wall geometries. In an SA DCR, the characteristic compression space is the one-dimensional SW $OI$ with the PR $\mathcal{A}_{O}^{\text{D}}$. As a two-dimensional being, any VSD imposed into the DCR can not be inside SW $OI$, i.e., separation point $S'$ and reattachment point $R'$ must be upstream and downstream of point $O$, respectively. On the other hand, the curved wall $\overline{MN}$ of CCR spreads out the SW into a series of CWs, which changes the original one-dimensional characteristic compression space into two-dimensional, and makes it possible for a suitable VSD to be imposed inside the space, as shown in Fig. \ref{subfig:VSD_CCR_SA}. For this reason, any VSD imposed into DCR's SA must be suppressed, while some VSDs imposed into CCR's SA satisfying subcase (i-iii) in relations (\ref{eq:Bistability_CCR}) will be stimulated.
\vspace{-5.0mm}
\section{Conclusions}
\label{sec:Conclusions}
\par The CCR flows' bistability is conjectured through a thought experiment, verified using 3D DNSs, and demonstrated with the VSD method. The essence of the bistability is that the curved wall geometry expands the characteristic compression space from one-dimensional SW into two-dimensional CWs, making both suppression and stimulation of VSD possible. However, as an intrinsic property, CCR's bistability does not originate from VSD, but from the bifurcation characteristics of Navier-Stokes equations. Thus, its specific presentation process could be diverse, corresponding to the hystereses induced by different parameter variations, and some 2D cases have been reported \citep{hu2020bistable,zhou2021mechanism}.
\par We have to say that there are some difficulties in verifying CCR's bistability in wind tunnels with conventional experimental methods, whose hidden premise is that the test flow's stable state is unique. Usually, the fixed flow parameters and wall geometries make the other potential state hard to appear. The demonstration in this paper could increase experimental aerodynamicists' confidence in observing CCR's bistable states in wind tunnels through ingenious methods. After all, it took more than 100 years from Mach's discovery \citep{mach1878uber} of two SR patterns to the observation of SR's bistable states in the wind tunnel \citep{chpoun1995reconsideration}. In some sense, the two geometrical parameters, curvature radius $R$ and wedge angle $\theta_{w}$, play the same role in the bistabilities of CCR flows and SRs, respectively.
\par SBLIs, represented by CCR flows, and Shock-Shock interactions (SSIs), represented by SW reflections \citep{ben2001hysteresis}, often dominate the complex flow in supersonic/hypersonic flight together. Therefore, more complex multistable shock patterns will be formed when multistable states of SBLIs and SSIs interact with each other, and the resultant complex aerothermodynamic characteristics need to be paid more attention in the future. 
\vspace{-5.0mm}
\section*{Acknowledgment}
We are grateful to Prof. Tao Zhu of LAMFR for his helpful discussion about the feasibility of wind tunnel experiments. This work was supported by the National Key R \& D Program of China (Grant No.2019YFA0405300). We look forward to receiving helpful and valuable comments from reviewers.
\vspace{2.5mm}
\par \hspace{-4.5mm} \textbf{Declaration of interests.} The authors report no conflict of interest.
\vspace{-2.5mm}
\bibliographystyle{jfm}
\bibliography{jfm2}


\end{document}